\documentclass[12pt, a4paper]{article}
\usepackage[font=footnotesize]{caption}
\usepackage[utf8]{inputenc}
\usepackage{graphicx}
\usepackage{authblk}
\usepackage{booktabs}
\usepackage{graphicx}
\usepackage{verbatim}
\usepackage{url}
\usepackage{caption}
\usepackage{subcaption}
\usepackage{indentfirst}
\usepackage{url}
\usepackage{hyperref}
\title{City Motifs as Revealed by Similarity Between Hierarchical Features}
\author{Guilherme S. Domingues \and Eric K. Tokuda \and Luciano da F. Costa}

\date{
São Carlos Institute of Physics - DFCM \protect\\
University of São Paulo \protect\\
P.O. Box 369, São Carlos, S.P. \protect\\
13560-970 Brazil
}
\hypersetup{draft}
\begin{document}
\maketitle

\begin{abstract}
Several natural and theoretical networks can be broken down into smaller
portions, or subgraphs corresponding to neighborhoods.  The more frequent of these neighborhoods can then be understood as motifs of the network, being therefore important for better characterizing and understanding of the overall structure.   Several developments in network science have relied on this interesting concept, with ample applications in areas including systems biology, computational neuroscience, economy and ecology. The present work aims at reporting an unsupervised methodology capable of identifying motifs respective to streets networks, the latter corresponding to  graphs obtained from city plans by considering street junctions and terminations as nodes while the links are defined by the streets.  Remarkable results are described, including the identification of nine stable and informative motifs, which have been allowed by three critically important factors: (i) adoption of five hierarchical measurements to locally characterize the neighborhoods of nodes in the streets networks; (ii) adoption of an effective coincidence methodology for translating datasets into networks; and (iii) definition of the motifs in statistical terms by using community finding methodology.  The nine identified motifs are characterized and discussed from several perspective, including their mutual similarity, visualization, histograms of measurements, and geographical adjacency in the original cities.  Also presented is the analysis of the effect of the adopted features on the obtained networks as well as a simple supervised learning method capable of assigning reference motifs to cities.
\end{abstract}

\section{Introduction}

Through a long period of time, cities unfolded as a means to provide resources to humans, including basic infrastructure as well as access to resources such as transportation, food, health, leisure, etc.  At the same time, city planning has had to adapt effectively to environmental constraints, including geographical and climatic characteristics.  Each city can thus be understood as a solution to the specific demands and constraints at varying levels of optimization.

Given that the spatial and topological organization underlying resulting cities are close and directly related to the above observed aspects, their respective study
(e.g.~\cite{rosvall2005networks,strano2013urban,porta2006network,buhl2006topological,louf2014typology,barthelemy2016structure,batty1994fractal,batty2007cities,batty2013new}) provides valuable means not only for better understanding how cities are organized, but also for possibly identifying how specific topological features of a city may be related to urbanistic and transportation aspects.  Respectively obtained results and insights can then be shared as part of planning and improvement approaches.

While the overall topology of a whole city can be characterized in terms of overall respective measurements, including average properties of blocks and streets, this type of global characterization cannot account for varying interconnectivity possibly taking place at different portions of the city.  For instance, even if a city is found to have blocks with an average of 4 sides, there may still be blocks with 3, 5 or more sides.  In addition, some portions of a city can be more or less densely covered by streets.  As a consequence, although global characterization of a city organization can provide valuable respective information, it is also of particular interest to perform studies of \emph{local} topological properties of cities, focusing on a size-limited neighborhood around each of the points of interest, which are henceforth understood as corresponding to every crossing between two or more streets or avenues.  The cities to be analysed are assumed to be represented as respective complex networks (e.g.~\cite{netwsci,newman,costa2011analyzing,boccaletti2006complex}), which can be achieved by representing streets crossings as nodes, while the streets or avenues between two nodes are taken as the network links.

\begin{figure}[ht]
    \centering
    \includegraphics[width=.5\textwidth]{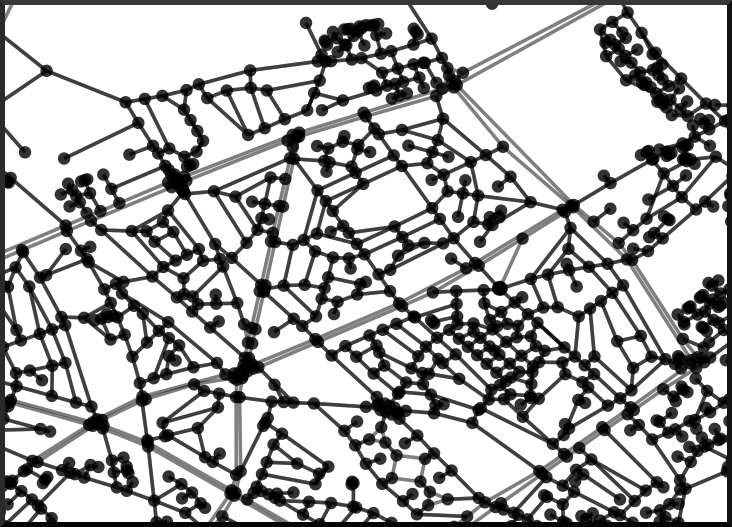} \vspace{0.5cm}
    \caption{A small portion of Liverpool, UK, illustrating the potential diversity of patches, or neighborhoods around each node, characterized by diverse topological properties.  The identification of recurrent neighborhood types, here called \emph{city motifs}, provides subsidies for developing several analysis aimed and better understanding and optimizing cities.  The identification of city motifs constitute the main objective of the present work.}
    \label{fig:crop}
\end{figure}

Figure~\ref{fig:crop} illustrates a small portion of a city (Liverpool,UK) involving several distinct types of neighborhoods with varying local properties, including highly regular square blocks, less regular regions, as well as streets dead ends.  The identification of the recurrent neighborhood types, or motifs, could contribute to developing and applying enhanced approaches not only to the characterization of cities, but also their better understanding, planning and optimization.

In this work, the neighborhood of a node is understood to incorporate all nodes that are taken into account by the adopted measurements.  Therefore, each of these neighborhoods will be henceforth specified in terms of the node (street crossings or termination point) to which it refers to, henceforth called \emph{reference node}, as well as to the number of hierarchical levels considered around that node.  The neighborhood of a node $i$ considering $H$ hierarchical levels will be expressed as $\eta_H(i)$.  It should be kept in mind that, henceforth in this work, the term \emph{neighborhood} will refer to the small subgraph around a given reference node, and not to the concept of city neighborhoods meaning a district within a city.

Local characterization of cities in terms of neighborhoods paves the way to a particularly interesting perspective, namely trying to identify common, frequent and recurrent local patterns of interconnectivity.  Indeed, the present work suggests a procedure for identifying and characterizing recurrent neighborhood topologies, which will be called \emph{mofits}, of a given city in terms of a respective complex network.

The concept of motifs in networks (e.g.~\cite{milo2002,budach2018}) has allowed several interesting results in network science, with ample applications in biochemistry, neurobiology, ecology, engineering, economy~\cite{xie2021}, transportation and infrastructure~\cite{jin2019}. Because of the intrinsic small topological variations expected to be found in city networks, the identification of possible motifs needs to be done statistically (e.g.~\cite{nissen2001rna,d2006dna}) while taking into account a set of informative local topological measurements.  It could be expected that highly regular, orthogonal neighborhoods, in terms of its nodes connections, would produce a clearly defined cluster, having an associated motif of orthogonal positioning, being very frequent in the city and with characteristic properties, such as constant degree distribution and a small clustering coefficient.  Other possibly expected motif would be the dead ended streets (degree one) and triangular blocks.

In the present work, we perform unsupervised identification of the motifs by using the coincidence methodology~\cite{costa2021coincidence,costa2021caleidoscope,costa2021onsimilarity}, which maps the neighborhoods into a respective network, so that the motifs are associated to respectively identified communities or modules.

Consisting of a combination of the widely employed Jaccard similarity index, adapted to real values~\cite{costa2021further,costa2021multiset,costa2021onsimilarity}, and the interiority (or overlap)  index (e.g.~\cite{vijaymeena2016survey}), the coincidence similarity provides an effective means for translating datasets, with entries characterized by respective measurements or features, into respective networks or graphs~\cite{costa2021coincidence,costa2021caleidoscope}.  In addition to its potential for obtaining particularly detailed and modular networks~\cite{costa2021elementary}, the coincidence methodology can also incorporated a parameter, namely $0 \leq \alpha \leq 1$, that allows the control of the relative contribution of positive and negative features pairwise relationships on the resulting similarity index~\cite{costa2021onsimilarity,costa2021coincidence}. For instance, by making $\alpha < 0.5$, the influence of positive joint variations can be attenuated so that a more detailed and modular pattern of interconnectivity can often be obtained, that could not be achieved by using the standard coincidence index (with $\alpha=0.5$).

The proposed methodology starts with a given city being represented in terms of its street network, in which nodes correspond to crossings between two or more streets and to terminations of streets, while the streets themselves give rise to the respective links.  A neighborhood with a specific extension $H$ is then obtained around each of the streets network nodes, and respective topological measurements are obtained.  Given that we are interested in studying varying neighborhood extensions $H$, node-centered topological measurements become of particular relevance for the characterization of the topological properties within the $H-$neighborhood around each of the streets network nodes. The coincidence methodology is then applied between the features of each possible pairwise combination of the $H-$neighborhood of the streets network nodes, therefore yielding a new network that, though with the same number of nodes, presents links whose strengths correspond to the similarity between the features of all possible pairs of $H-$neighborhoods.   This network is henceforth called the \emph{neighborhood network -- NN}.

As a consequence of the above described approach, two nodes in an NN will be strongly interconnected whenever the neighborhoods associated to those nodes have strongly similar local topological properties.  Several interesting information and insights can be potentially obtained from these networks.  For instance, a narrow distribution of interconnection strengths will indicate that most of the node neighborhoods are similar, while wider strength distributions will reveal that the neighborhoods of the specific city of interest are noticeably heterogeneous.  In addition, in case the obtained NN presents a well-defined modularity, community detection methods (e.g.~\cite{newman2006modularity,fortunato2010community,reichardt2006statistical,fortunato2016community}) can be applied in order to identify the main modules, each of which will indicate a mesoscopic region of the city presenting particular topological properties.

Each of the communities identified in NNs will constitute a candidate for a topological \emph{motif}.  Therefore, in addition to studying the similarity between the topological properties of node neighborhoods across varying topological scales, the present work also aims at investigating if the motifs identified among two or more cities can be inter-related. For instance, one such module recurring between several cities can be understood as a possible shared motif.  In order to develop these studies in a systematic manner, we apply the coincidence methodology to three Brazilian cities, identify the respective modules, and then apply the coincidence approach to derive a network of these modules, which will be henceforth called the \emph{motif network} of a set of cities.  The communities eventually identified in this network will therefore provide indication about shared motifs of neighborhood topology.
Observe that a motif network constitutes of a neighborhood network to which specific types of motifs have been assigned.

The obtained results revealed a surprising level of consistency and stability of nine identified motifs, which have specific visual, topological, cross-similarity, and adjacency properties, all of which having being quantified in an objective manner in the present work.  

In order to complement the discussion of the obtained results, we also performed an analysis of the influence of the adopted five hierarchical measurements on the resulting neighborhood networks, which was developed by using an approach that is also based on the coincidence methodology~\cite{costa2021elementary,costa2022cities}.

The observed generality of the identified motifs, as well as their dependence on local neighborhoods in the respective streets networks, motivated the proposal of a simple supervised method for assigning motifs to cities.  This method, which is described and illustrated in the present work, involves using the instances of motifs identified for the reference cities as a reference table, so that neighborhoods of other cities can be assigned by taking into account the motif type corresponding to the smallest distance between the local hierarchical measurements of the entries in the reference table and the neighborhoods to be classified.  The method was shown to perform remarkably well for the case of a forth Brazilian city, namely Birigui.

This work starts by providing a non-exhaustive review of related works and follows by presenting the data, basic concepts and methods, including hierarchical measurements, the coincidence methodology, and motifs identification.  The results are then presented and discussed respectively to features interrelationship, motifs characterization, and application to three Brazilian cities.  The effect of the adopted features on the respectively obtained networks is also addressed, and a simple procedure for assigning the nine types of identified motifs to generic cities is also presented.

\section{Related Works}

This section revises, a non-fully comprehensive manner, some of the works related to the main aspects and concepts developed in the current article.  

The comparison of networks, a topic of significant interest in network science, can be implemented based on different criteria, such as the network type, the degree distribution, and the presence of communities. In~\cite{choi2006comparing} the similarity between the internet backbone and air transportation network is addressed by considering the hierarchy and pattern of connections among world cities.  In~\cite{liao2015reconstructing}, four different standard similarity metrics (Common Neighbors, Jaccard, Resource Allocation and Leicht-Holme-Newman) are used to evaluate node similarity and reconstruct propagation networks based on the epidemics spreading dynamics. It is observed that temporal information can play a pivotal role on the reconstruction. In~\cite{cardillo2006structural}, different samples of the street networks of 20 different world cities are compared with basis on a set of measurements of spatial graphs, namely the meshedness, the number of short cycles of sizes three, four and five edges, the global efficiently and the cost. In particular, similarity is estimated between self-organized and planned cities.

The characterization of networks can be made locally. In the context of spatial networks, in ~\cite{hipp2012measuring} the authors propose defining neighborhoods based on social ties as well as on physical distance. They propose four alternative manners of doing, which are applied to data of students from North Carolina schools. In~\cite{li2007network}, in the scenario where connections are susceptible to noise, the authors consider a neighborhood scheme based on shared neighbors.

Networks can also be characterized in terms of the presence of pre-defined patterns, commonly known as \emph{motifs} (e.g.~\cite{alon2007network,milo2002network,stone2019network}). For instance, the distribution of triangles along a network has been used as an indicative of the tendency of the network to form clusters~\cite{costa2007characterization}.  The distribution of motifs has also been studied respectively to its effects on specific types of dynamics on networks~\cite{lodato2007synchronization,ciriello2008review,sporns2004motifs,liu2022temporal}.

There is a variety of reported applications of network motifs. In~\cite{boas2008chain,bordertrees}, the authors study network connectivity in terms of specific types of motifs: vertices connected in a sequential way such that the inner vertices have degree equal to two. They observe highly different distributions of these motifs between real-world and artificial networks.  In~\cite{larock2021sequential}, the authors analyze the distribution of motifs in directed networks, which they call sequential motifs. They propose a connection between sequential motifs and higher order networks, and analyze data from passenger trips through the airport network in the United States and also  article navigation in Wikipedia. Motifs have also been used to analyze data from mobile phone communication networks and related data, which can be used to study communication and human mobility patterns~\cite{schneider2013unravelling,stoica2009structure}.
Mobility patterns from tourists are studied in~\cite{yang2017quantifying}. The authors considered in their analysis temporal information (such as \emph{when} the places were visited and semantic information (the attractions). The temporal travel motifs in this case revealed popular duration of stays in each attraction while the topological motifs the frequent travel sequences among the attractions.

One particular application of motifs has been in the study of street networks~\cite{tsiotas2017topology,ping2006topological}. In~\cite{tsiotas2017topology}, the authors analyze how socioeconomic aspects of a city -- such as mobility, market and population -- associate to city street network patterns. They considered Greek cities and observed three distinct patterns: considering the central nodes, ring nodes and the mixture of the two.
In~\cite{ping2006topological} the authors study the frequency of motifs in public transportation networks in large Chinese cities.  One of the main findings regards the distribution of certain 3-node motifs, which seemed to be associated with the efficiency of the transportation system and robustness to failures.

\section{Materials and Methods}

\subsection{Streets Networks}

We considered a set of three Brazilian cities -- namely S\~ao Carlos, Lages and Imperatriz, with population between $100.000$ and $300.000$ inhabitants, whose streets were represented by complex networks, with each node representing streets crossing or termination, while the corresponding street as link between that pair of nodes.  The three considered cities are suitably located in distinct regions of Brazil, namely  north (Imperatriz), center (S\~ao Carlos), and south (Lages), therefore contributing to generality of the results. All these three cities have similar populations and are located in main land. The data for node localization and connections was obtained from the OpenStreetMaps~\cite{OSM} database. After obtaining the city networks,  hierarchical measurements were calculated for each node and then used for identification of possible motifs characterizing the cities topological organization.

\subsection{Hierarchical Measurements}
\label{subsec:hierarchical}

In this work, we considered a set of five hierarchical measurements (e.g.~\cite{travenccolo2008hierarchical,da2006hierarchical,ahnert2009connectivity}) for characterizing the node neighborhoods, to be taken as features in the coincidence methodology, as described in Section~\ref{sec:Coincidence}. The adoption of a $H-$neighborhood around the reference node $i$ implies the hierarchical measurements to be calculated relatively to the hierarchical levels $R_h(i)$ with $h = H-1$. The following measurements have been henceforth considered:

\begin{table}[!hp]
    \centering
    \begin{tabular}{l|c}
        Measurement & Symbol \\ 
        \hline
         Hierarchical degree & $hd$  \\
         Hierarchical clustering coefficient & $hc$ \\
         Convergence ratio & $cr$ \\
         Hierarchical number of nodes & $hn$ \\
         Hierarchical number of edges & $he$ 
    \end{tabular}
    \caption{Table of the measurements used in this work with respectively adopted symbols.}
    \label{tab:measurements}
\end{table}

\noindent \textbf{Hierarchical degree} ($hd$): 
The hierarchical degree $hd_h(i)$ of node $i$ at distance $h$ is defined as the number of edges between the hierarchical levels $R_h(i)$ and $R_{h+1}(i)$

\noindent \textbf{Hierarchical clustering coefficient} ($hc$):  Hierarchical clustering coefficient of node $i$ at distance $h$ is defined as
\begin{equation}
    hc_h(i) = 2 \frac{e_h(i)}{n_h(i)(n_h(i) -1)}
\end{equation}
where $e_h(i)$ is the number of edges connecting nodes of the hierarchical level $R_h(i)$ and $n_h(i)$ is the number of nodes of that hierarchical level.

\noindent \textbf{Convergence ratio} ($cr$): The Convergence ratio of node $i$ at hierarchical level $h$ is defined as the ratio between $hd_h(i)$ and the number of nodes in the next hierarchical level, i.e.
\begin{equation}
    cr_h(i) = \frac{hd_h(i)}{n_{h+1}(i)}
\end{equation}

\noindent \textbf{Hierarchical number of nodes} ($hn$):  The hierarchical number of nodes $N_h(i)$ in the hierarchical level $R_h(i)$ is defined as the number of nodes $n_h(i)$ inside $R_h(i)$, or the size of $R_h(i)$.

\noindent \textbf{Hierarchical number of edges} ($he$):  The hierarchical number of edges $he_h(i)$ among the nodes in the hierarchical level $R_h(i)$ is defined as the number of edges $e_h(i)$ between the nodes of $R_h(i)$ without considering edges connecting nodes of $R_{h+1}(i)$ or $R_{h-1}(i)$.

\subsection{The Coincidence Methodology}
\label{sec:Coincidence}

Several similarity indices have been considered respectively to diverse types of data and applications (e.g.~\cite{vijaymeena2016survey,mirkin1996mathematical,dataclust2012,akbas2014l1,costa2021onsimilarity}), including cosine similarity, correlation, and the Jaccard index.
Though the \emph{Jaccard index} (e.g.~\cite{costa2021onsimilarity,vijaymeena2016survey,wikijaccard}) has been extensively employed as a means of quantifying the \emph{similarity} between two sets, these applications have been mostly limited to categorical or binary data.  In addition, the Jaccard index has been shown not to be able to take into account how much the two compared sets are mutually internal one another~\cite{costa2021further}.  This motivated the consideration of the \emph{coincidence similarity index}~\cite{costa2021further}, corresponding to the product of the Jaccard index and the respective \emph{interiority} or \emph{overlap} index (e.g.~\cite{vijaymeena2016survey}).

By extending multisets (e.g.~\cite{Singh,Thangavelu,Knuth,heinz:2011,Blizard,Blizard2}) to real-valued data~\cite{costa2021multiset}, it was possible to derive a respective coincidence index that can be employed as a means to quantify the similarity between two real-valued vectors or even functions.  In addition, it has been shown that the Jaccard index can be decomposed into two major terms, one corresponding to the positive pairwise alignment of the signs of the compared values, and another to the anti-aligned pairs.  The linear combination of these two terms, respectively weighted by $\alpha$ and $1-\alpha$, yields the parametric coincidence index expressed as:

\begin{equation}
  \mathcal{C}_R(\vec{f},\vec{g},\alpha) = \mathcal{I}_R(\vec{f},\vec{g}) \  \mathcal{J}_R(\vec{f},\vec{g}, \alpha)
\end{equation}

where:

\begin{equation}
  \mathcal{J}_R(\vec{f},\vec{g},\alpha) =
    \frac{\sum_i \alpha   |s_{f_i} + s_{g_i}| \min\left\{ | f_i |, | g_i |\right\} 
  - (1 - \alpha)  |s_{f_i} - s_{g_i}| \min\left\{ | f_i |, | g_i |\right\} }
  {\sum_i \max\left\{ | f_i |, | g_i |\right\}}
\end{equation}

and: 
\begin{equation}
  \mathcal{I}_R(\vec{f},\vec{g}) = \frac{ \sum_i \min\left\{ | f_i |, | g_i |\right\}}
  {\min\left\{ \sum_i | f_i |, \sum_i | g_i | \right\}}
\end{equation}

We also have that $-2(1-\alpha) \leq J_R(\vec{x}, \vec{y}, \alpha) \leq 2\alpha$.

Thus, the parameter $\alpha$ allows an effective control of how the aligned and anti-aligned pairwise measurements are combined into the resulting overall coincidence value.  

In particular, when $\alpha=0.5$, the above index becomes identical 
to the product between the real-valued, parameterless Jaccard index and the interiority index, i.e.:
\begin{equation}
    \mathcal{C}_R(\vec{x},\vec{y}, \alpha=0.5) = 
     \mathcal{J}_R(\vec{x},\vec{y}) \mathcal{I}_R(\vec{x},\vec{y}) 
\end{equation}

The real-valued coincidence index has been applied~\cite{costa2021coincidence,costa2021caleidoscope} to translate datasets, with each data element characterized in terms of $M$ measurements or \emph{features}, into respectives graph or networks whose interconnecting weights between each two nodes correspond to the respective coincidence values between the features of those two nodes.  These coincidence networks can be then thresholded by $T$ so as to yield networks with weights limited to $0$ and $1$.  It is also possible to preserve the values of the coincidences above $T$.

It has been shown~\cite{costa2021coincidence,costa2021caleidoscope,costa2021elementary} that the interconnectivity of the resulting coincidence networks strongly depends on the values set for $\alpha$, in the sense that higher values of $\alpha$ will imply more intensely interconnected networks.  However, these networks tend to become too interconnected, to the point that the respective interconnection details and modularity are severely blurred and cluttered.  This is precisely where reductions of the parameter $\alpha$ can substantially contribute to limiting the overall connectivity, contributing to obtaining more detailed and modular networks.  Indeed, it has been verified~\cite{costa2021coincidence,costa2021elementary} that the modularity of the coincidence networks tends to substantially increase as $\alpha$ is reduced.

In this manner, the coincidence methodology for quantifying similarity between real-valued vectors and functions (as well as other types of data) incorporates several interesting features deriving from the  Jaccard and interiority indices combined with the critically important control of the resulting overall interconnectivity by varying the parameters $\alpha$.

\begin{figure}[ht]
    \centering
    \includegraphics[width=\textwidth]{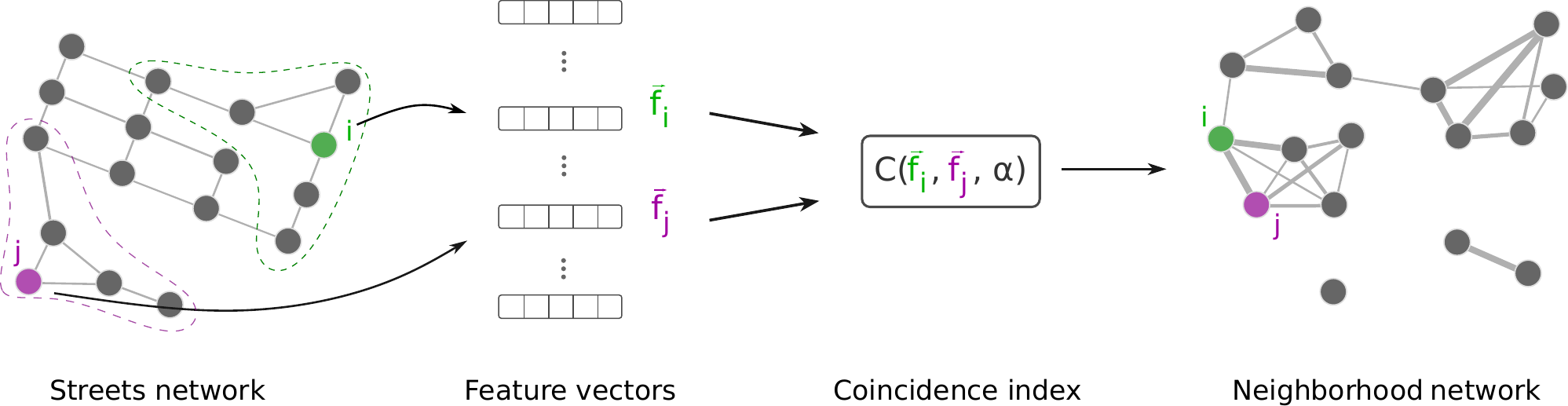}
    \caption{Diagram illustrating how a streets network is translated into the respective coincidence network.  The neighborhood around each node
    (e.g.~$i$ and $j$), is identified and the respective hierarchical measurements are calculated.  The example in this figure considers $h=1$, therefore including only its first neighbors.  The coincidences between each resulting feature vectors are calculated and taken as the weights between the nodes in the coincidence network.   Observe that both the streets and coincidence network have the same number of nodes.}
    \label{fig:pipeline}
\end{figure}

In the current work, for each city, as illustrated in Figure~\ref{fig:pipeline}, the neighborhood around each node $i$ is identified and the respective hierarchical measurements obtained and organized into a respective feature vector as follows:
\begin{equation}
    \vec{f}_i = [ hd(i), hc(i), cr(i), hn(i), he(i)]
\end{equation}

The obtained features are then supplied to the above described coincidence methodology  in order to deriving respective coincidence networks for each individual city.  Observe that each neighborhood in the streets network becomes represented by a single node in the neighborhood network, corresponding to the respective neighborhood reference node.  As a consequence, two adjacent nodes in the latter network will necessarily imply some overlap between their respective neighborhoods in the streets network.

Similarity is intrinsically related to connectivity (e.g.~\cite{comin2020complex}), providing a means for obtaining complex networks  (e.g.~\cite{costa2004complex,onnela2004clustering,yang2021node,putra2017evaluating,backes2010shape}).  The coincidence similarity, which has been applied as a means of translating datasets into respective complex networks~\cite{costa2021coincidence,costa2021caleidoscope} is adopted henceforth in the present work. More specifically, after being standardized, the features describing the dataset are taken into account while calculating the coincidence between every pair of data elements, resulting a coincidence network in which each node corresponds to a data element while the links are determined by the respective pairwise coincidence similarity indices.

The standarization of each of the adopted features $x_i$, respective to data elements $i$, can be implemented as follows:
\begin{equation}
    \tilde{x}_i = \frac{x_i - \mu_{x}} {\sigma_{x}}
\end{equation}

where $\mu_{x}$ and $\sigma_{x}$ are the average and standard deviation of feature $x_i$ taken along the whole considered dataset.

\subsection{Motifs Identification}

As observed in the introduction of the present work, given the diversity of interconnections typically observed in streets networks, the \emph{neighborhood motifs} (NMs) to be considered here have a statistical nature, in the sense that a given motif type can present intrinsic small topological variations.  The basic hypothesis of our approach regarding the NMs is that they have some level of generality and recurrence not only within a given city, but also across other cities.  Thus, the problem of motif identification as addressed in the present work can be stated as: given a set of cities and respective neighborhoods characterized by associated features, to find sets of these neighborhoods that are strongly similar one another while being distinct to the other neighborhoods.  

The resource to be applied in order to find these groups of similar neighborhoods, which will be taken as the NMs, consists of the application of the coincidence methodology~\cite{costa2021coincidence,costa2021caleidoscope,costa2021onsimilarity}.  More specifically, we estimate the coincidence similarity between each pair of neighborhoods obtained from all the adopted cities, and a single network is therefore obtained from each neighborhood while the coincidence similarity between each pair of nodes corresponds to the respective link weight.  In order to simplify the resulting network, its links with coincidence values smaller than a given reference $T$ are subsequently ignored, therefore yielding a weighted network (a binary network would be otherwise obtained by standard thresholding).  This operation is henceforth referred to as \emph{blanking}.  

The NMs can then be identified as being associated to the main detected communities having at least a minimum number of nodes $N$. The community detection is performed independently in the combined neighborhood network and also in each considered city neighborhood network as a means to identify the correspondence among the detected communities across cities.  This is implemented for each city at a time.  For each community $m$ in a given city, it is verified which among the communities in the combined network contains the largest number of the nodes in $m$, which is taken as the corresponding community. 

\begin{figure}[ht!]
    \centering
    \includegraphics[width=\textwidth]{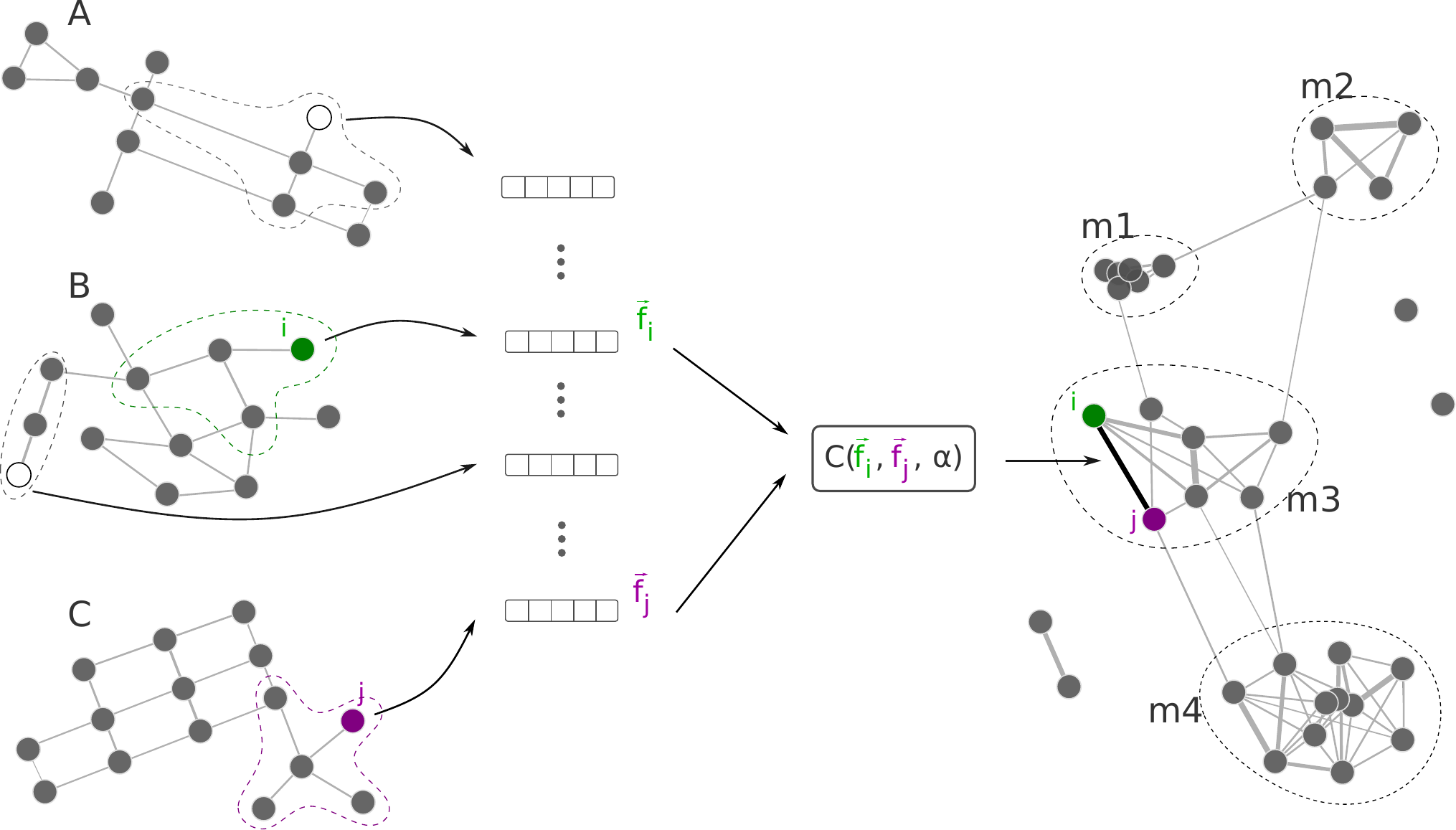}
    \caption{Given three generic cities ($A$, $B$, and $C$), all possible neighborhoods with hierarchical level $H$ are identified, and the respective standardized hierarchical measurements are organized as feature vectors, four of which are shown in the figure.  The coincidence similarity is then calculated between each pair of these feature vectors.  By blanking these values by $T$, the combined neighborhood network CN can then be obtained.  Each detected community in the latter network is then understood as a neighborhood motif -- NM.}
    \label{fig:allcities}
\end{figure}

The suggested methodology for identifying the NMs is illustrated in Figure~\ref{fig:allcities} respectively to three generic cities $A$, $B$, and $C$ of interest.  The adopted five hierarchical features are standardized (e.g.~\cite{gewers2021principal}) along all neighborhoods of a considered city before coincidence similarity estimation.

\section{Results and Discussion}

\subsection{Neighborhood Characterization}

\begin{figure}[ht!]
    \centering
    \includegraphics[width=\textwidth]{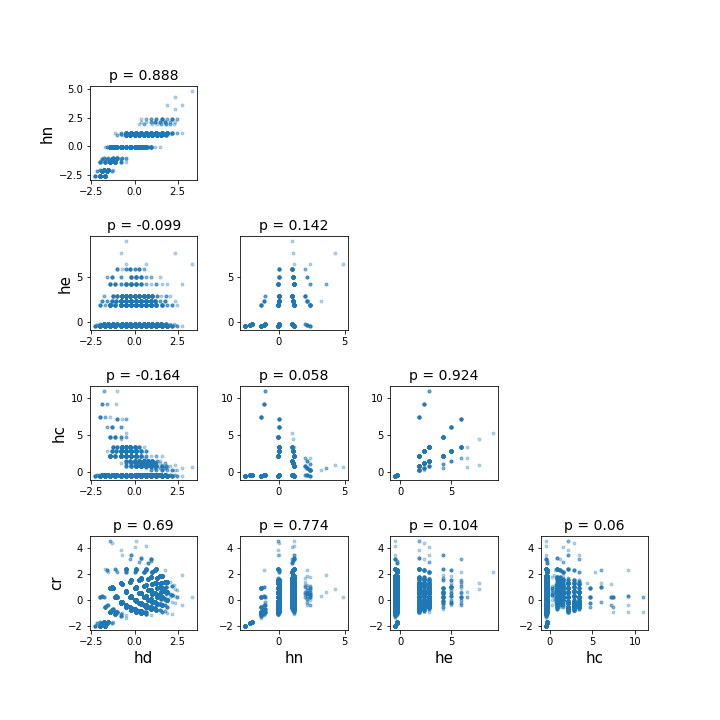}  \vspace{0.5cm}
    \caption{Scatterplots for each pair of adopted hierarchical measurement considering all the neighborhoods $\eta_2(i)$ in the cities of S{\~a}o Carlos, Lages and Imperatriz, including the respective Pearson correlation coefficient.}
    \label{fig:corr}
\end{figure}

As a first step in our approach, we calculated the five hierarchical measurements for each neighborhood $\eta_2(i)$ ($H=2$) respective to each node $i$, which were used to characterize locally the topological properties of the streets network.  Figure~\ref{fig:corr} presents the scatterplots obtained for each pair of hierarchical measurements considering the nodes for all the three considered cities.  

The obtained result is interesting because it indicates relatively minor pairwise correlations between several of the measurements.  More specifically, only four out of the ten possible pairwise features combinations resulted in high Pearson correlation coefficients, with the remainder six relationships being characterized by markedly small correlation values.  Most obtained correlation values are positive.  This corroborates that the five measurements are little redundant one another, therefore complementing one another with respect to the characterization of the topological properties within each considered neighborhood $\eta_2(i)$.

\subsection{Neighborhoods Networks}

\begin{figure}[ht!]
    \centering
    \includegraphics[width=0.2\textwidth]{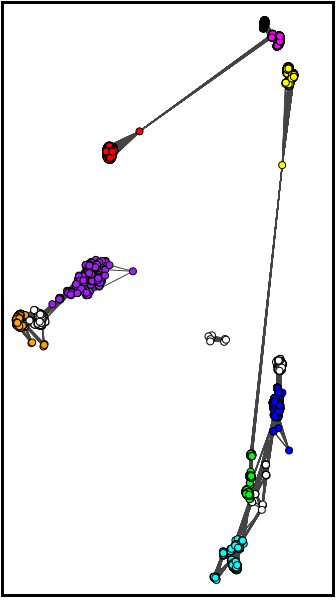} \hspace{0.5cm}
    \includegraphics[width=0.2\textwidth]{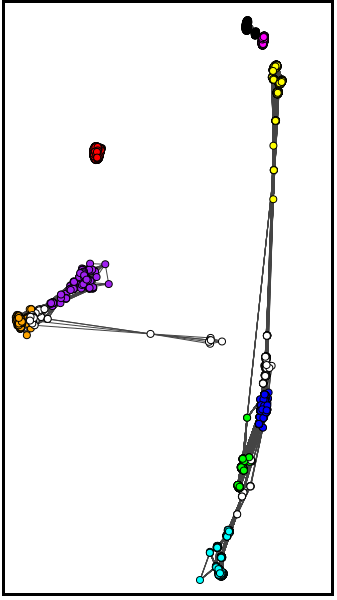} \hspace{0.5cm} 
    \includegraphics[width=0.2\textwidth]{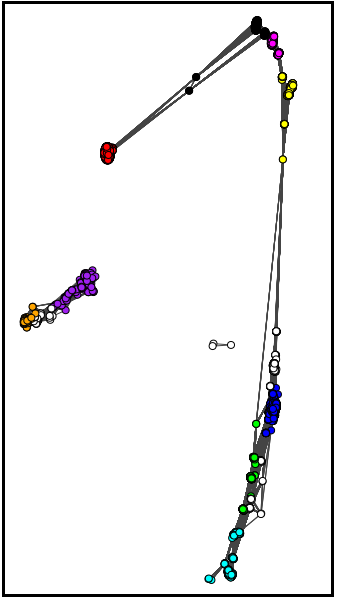} \hspace{0.5cm}
    \includegraphics[width=0.2\textwidth]{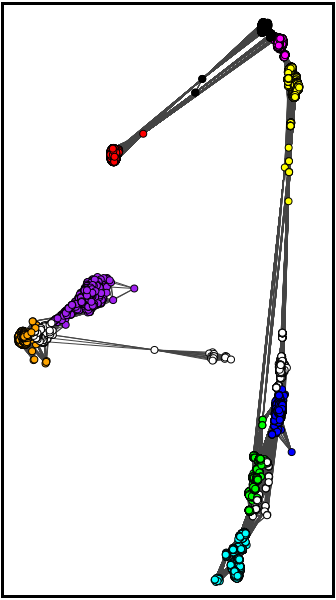} \\
    (a) \hspace{2.7cm} (b) \hspace{2.8cm} (c) \hspace{2.7cm} (d)
    \caption{The neighborhood networks obtained for the cities of S{\~a}o Carlos (a), Lages (b) and Imperatriz (c).  Each node $i$ corresponds to the respective  $\eta_2(i)$ neighborhood, while the link values are the respective pairwise coincidences assuming $T= 0.1$ and $\alpha = 0.1$, which a good separation between the nodes into clusters.  The respectively obtained combined network is shown in (d), integrating all the neighborhoods from the three considered cities. The colors indicated the nine identified motifs (Fig.~\ref{fig:motifs}), with the neighborhoods not associated to motifs being represented in white.  All networks have been visualized using the same layout as that obtained for the combined network.}
    \label{fig:NNs}
\end{figure}

Figures~\ref{fig:NNs}(a) to (c) present the neighborhood networks obtained respectively to the S{\~a}o Carlos, Lages, and Imperatriz cities, while Figure~\ref{fig:NNs}(d) shows the respectively obtained \emph{combined network} containing all neighborhoods from the three cities as well as their interrelationships. These visualizations were obtained by using the Fruchterman-Reingold~\cite{fruchterman1991graph} method.

Of particular interest is the relatively high modularity of all obtained neighborhood networks, which was mostly allowed by the strict similarity quantification implemented by the coincidence methodology, as well as the mutual coherence between the obtained mappings and communities.  The combined network in Figure~\ref{fig:NNs}(d) provides the basis for identifying the city motifs, which was done by detecting the respective communities using the Fruchterman-Reingold~\cite{fruchterman1991graph} method.  Each of the nine identified community with at least $N=700$ nodes (neighborhoods) were understood as putative motifs.

The obtained result is critically important because the neighborhoods from three distinct cities were found to be partitioned into communities that are mutually congruent, in the sense that the clusters of nodes respective to each of the nine identified communities are well separated in all the four obtained networks.  This indicates that the nine identified motifs tend to be self-consistent not only within the same city, but across the set of considered cities, therefore suggesting that these motifs could be universal among other cities.  This possibility is supported by the fact that the neighborhoods from which the motifs have been identified are strictly local, considering only $H=2$ neighborhood levels, implying that the adopted hierarchical measurements are largely resilient to border effects.

Table~\ref{tab:motifs} presents the number of motifs of each type, from 1 to 9, identified in the combined network.  The table presents the motif types in decreasing order of the most frequent type.

\begin{table}[]
    \centering
    \begin{tabular}{c|c|c|c}
        Motif identif. & Motif color & N. of nodes & Rel. Freq. \\ 
        \hline  
        $m1$ & blue & 4311 & 22.109 \% \\
        $m2$ & yellow & 2881 & 14.775 \% \\ 
        $m3$ & red & 2561 & 13.134 \% \\ 
        $m4$ & cyan & 2079 & 10.662 \% \\ 
        $m5$ & magenta & 1944 & 9.970 \%  \\ 
        $m6$ & green & 1436 & 7.364 \% \\ 
        $m7$ & orange & 1377 & 7.062 \% \\ 
        $m8$ & purple & 909 & 4.662 \% \\
        $m9$ & black & 760 & 3.898 \% \\ 
        Unassigned & white & 1241 & 6.364 \%
    \end{tabular}
    \caption{The identification and number of neighborhoods in the combined network corresponding to each of the nine identified city motifs.}
    \label{tab:motifs}
\end{table}

\subsection{City Motifs and Motifs Network} \label{sec:citymotifs}

The city motifs were identified from the combined motif network in Figure \ref{fig:motifs}.  More specifically, the community finding method \cite{kamada1989algorithm} was applied to that network, corresponding to possible respective city motifs obtained for the three considered cities.  

\begin{figure}[ht!]
    \centering
    \includegraphics[width=0.4\textwidth]{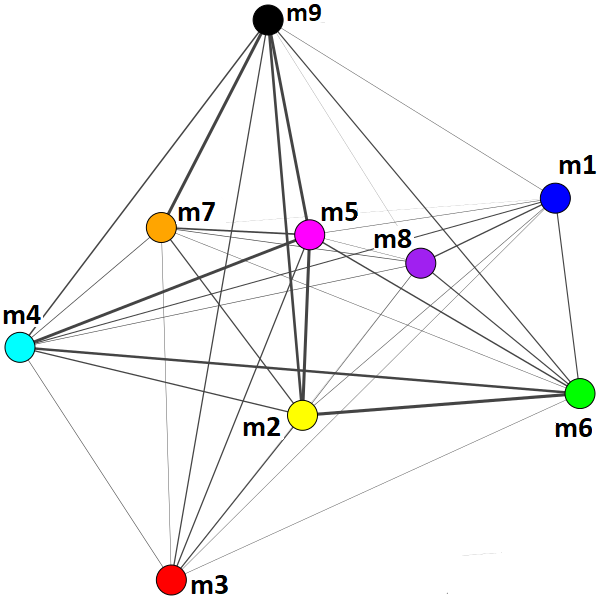} \vspace{0.5cm}
    \caption{Coincidence similarity network obtained for the nine identified city motifs.  Each node corresponds to one of the motifs, while the width of the links is proportional to the respective pairwise coincidence between the five features adopted for characterizing each neighborhood $\eta_2(i)$, assuming no blanking ($T = 0$) and $\alpha = 0.1$.}
    \label{fig:redeMotif}
\end{figure}

Figure~\ref{fig:redeMotif} illustrates the coincidence network obtained for the nine identified motifs.  Each node in this similarity network corresponds to one of the identified motifs, while the width of the link between two motifs $i$ and $j$ indicates value of the respective coincidence similarity between the respective features densities.  The node with the largest strength (sum of coincidences respective to its links) corresponds to the city motif 5, which can thus be understood as being more mutually similar to several of the remainder motifs.  Also worth noticing is the relatively stronger relationship between motifs 2-4-6, 2-5-9, 4-5, as well as 7-9, meaning that they are intrinsically more similar one another.  Motifs 8, 1, and 3 have the smallest coincidence strengths, being therefore relatively more distinct to the remainder motifs.

The motifs network in Figure~\ref{fig:redeMotif} also provides subsidies for further reducing, if necessary, the number of motifs, which can be done by merging pairs of strongly interconnected motifs.

\subsubsection{Motifs Characterization}

In this section we discuss the nine identified city motifs in terms of three particularly important respective perspectives: (i) visual appearance; (ii) histograms of features; and (iii) geographical adjacency between motifs.  

\begin{figure}[!htbp]
    \centering
    \includegraphics[width=0.95\textwidth]{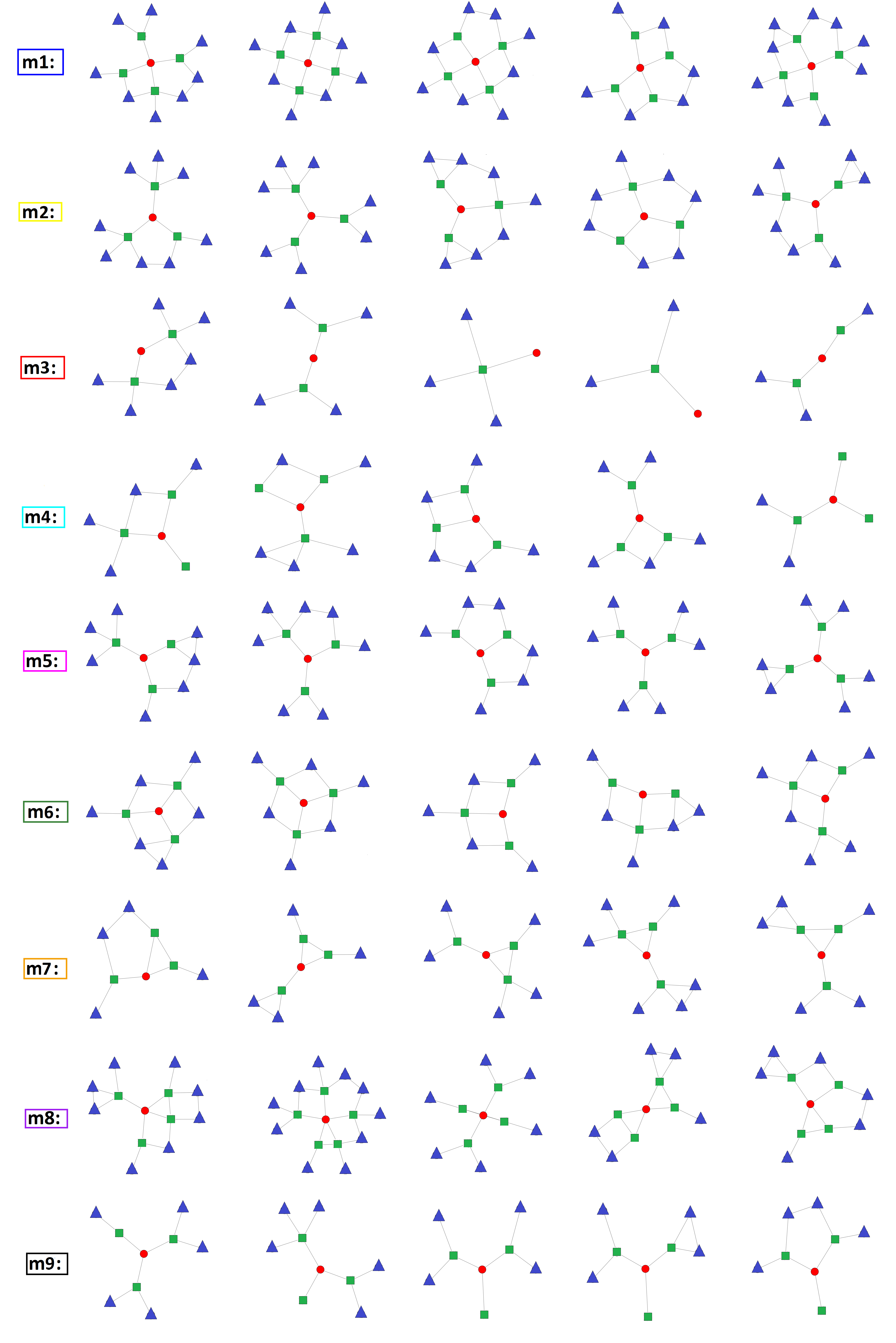}
    \caption{Examples of the nine motifs identified from the combined neighborhoods netwoks considering the Brazilian cities of S{\~a}o Carlos, Lages and Imperatriz, with the reference nodes shown as circles, while the respective first and second neighborhoods are depicted as squares and triangles, respectively.  The motifs (rows) are presented in decreasing order of respective frequency in the three considered cities.  As expected, small variations can be observed among motifs from the same type, which justifies the adopted statistical approach for motif identification.   The illustrated nine motifs were identified from a coincidence networks derived from neighborhoods described by the five adopted hierarchical measurements.}
    \label{fig:motifs}
\end{figure}

Figure~\ref{fig:motifs} depicts five samples of motifs of each of the nine identified types.  These samples correspond to those with highest strength in the combined network within each detected motif. The reference node has been marked in red, while its first and second neighborhoods are marked in green and blue, respectively.

Of particular relevance is the high level of similarity observed among samples from the same type of city motif.  In addition, despite intrinsic statistical variations, the samples from distinct motifs resulted with marked topological differences.  For instance, motifs $m1$ and $m8$ are characterized by reference nodes with degrees equal to four, in contrast to degrees one or two observed for the reference nodes of $m3$.   Motifs $m2$, $m4$, $m5$, $m6$, $m7$, and $m_9$ have reference nodes with degree equal to three.   The distinction between the motifs having the same reference node degree is accounted for by the other considered hierarchical features, which cannot be straightforwardly discerned by visual analysis.  

\begin{figure}[ht!]
    \centering
    \includegraphics[width=0.97\textwidth]{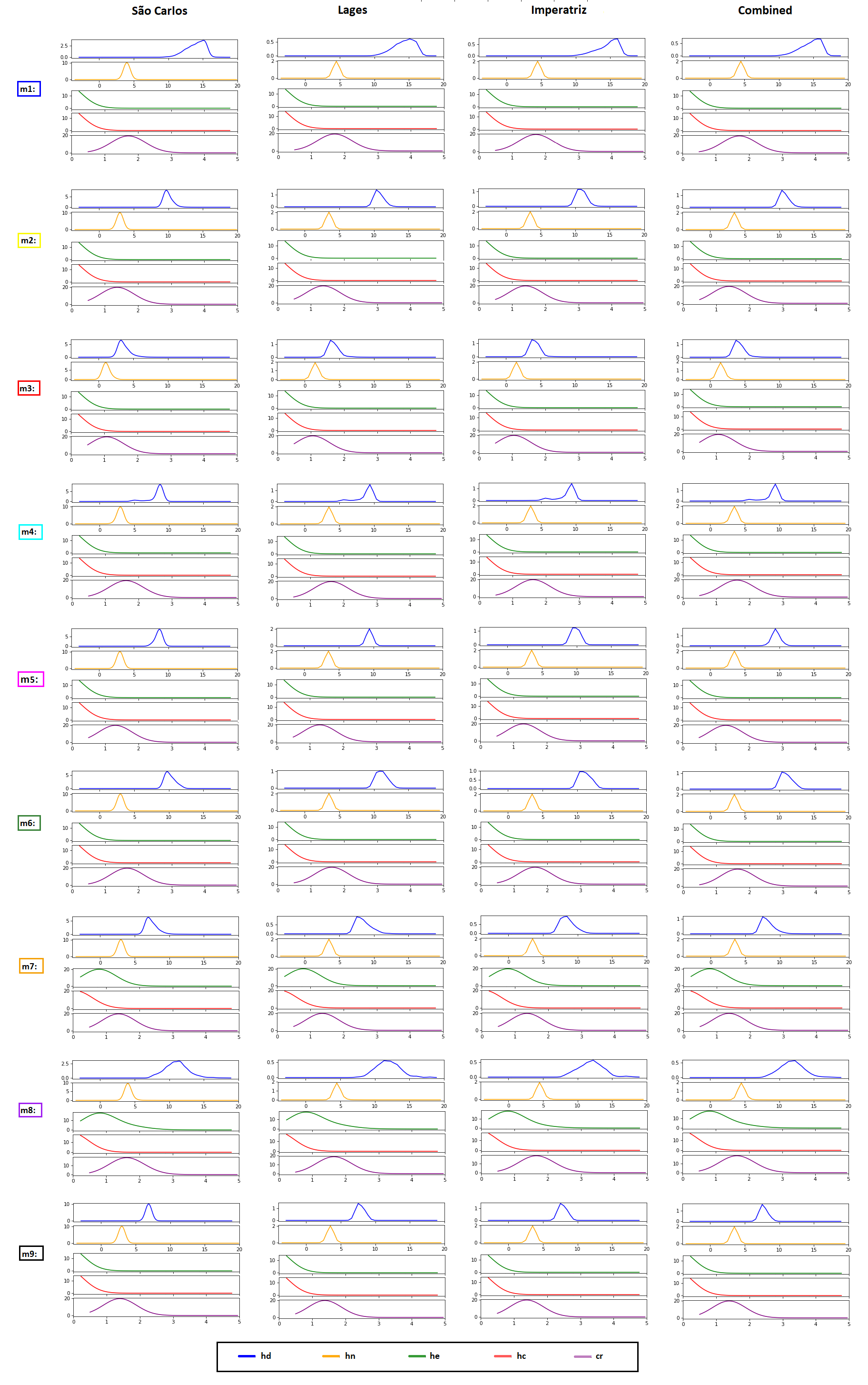}
    \caption{Relative frequency histograms of the five adopted hierarchical measurements obtained for the nine identified city motifs respectively to the S{\~a}o Carlos, Lages, Imperatriz, and combined neighborhood networks.  Of particular interest is the high similarity level between the shapes of the histograms respective to the same city motif type, while the obtained measurements vary among distinct motif types.}
    \label{fig:hist}
\end{figure}

In order to characterize the identified motifs in a more comprehensive manner, it is necessary to resource to the histograms of the adopted five hierarchical measurements obtained for each of the nine identified motifs respectively to the S{\~a}o Carlos, Lages, Imperatriz, and combined neighborhood networks, which are presented in Figure~\ref{fig:hist}(a) to (d), respectively.

These histograms provide an objective characterization of the nine identified motifs, therefore complementing the preliminary visual analysis of the motifs. Of particular importance is the fact that the histograms obtained for each of the four neighborhood networks have similar shapes, corroborating the consistency and generality of the identified motifs.  At the same time, markedly distinct histogram shapes can be observed between two distinct motif types.  To a considerable extension, these important results have been allowed by the choice not only of the informative hierarchical measurements, but also by the strict similarity characterization implemented by the coincidence methodology.

Several types of distinctions can be discerned among different motif types.  For instance, the histogram of \emph{hierarchical degree} $hd$ tends to appear at different positions along the horizontal axes and with varying shapes.  Consequently, this feature can be deemed to be of particular importance for distinguishing between the nine identified city motif types.  In addition, observe that the histogram of the hierarchical number of edges resulted null or nearly null for several motif types.  Another factor contributing to the differentiation between the identified city motifs consists of the positions of the histograms obtained for the other measurements.

A more comprehensive characterization of the identified motifs taking into account the distributions of all the five adopted hierarchical features is presented in Section~\ref{sec:motifs}.

\begin{figure}[ht!]
    \centering
    \includegraphics[width=.25\textwidth]{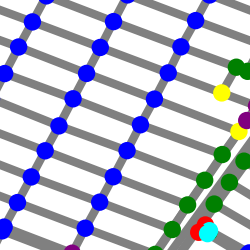} \qquad
    \includegraphics[width=.25\textwidth]{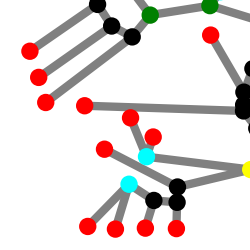} \qquad
    \includegraphics[width=.25\textwidth]{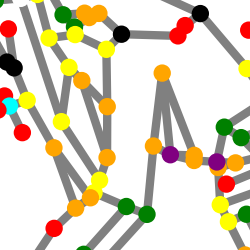} \qquad \\ (a) \hspace{3.5cm} (b) \hspace{3.5cm} (c)
    \caption{Visualization of instances of the identified motifs in the streets networks of the analyzed cities. (a): $m1$ (blue) nodes appear as representing highly regular and rectangular blocks, with predominance of $m6$ (green) on their borders. (b): $m3$ (red) are end nodes, generally linked to the network through $m4$ (cyan) and $m9$ (black). (c): the occurrence of $m7$ (orange) and $m8$ (purple), as part of triangular blocks.}
    \label{fig:zooms}
\end{figure}

Another important property of the city motifs concerns their geographical adjacency in the original streets network (see Figure \ref{fig:zooms}).  

\begin{figure}[ht!]
    \centering
    \includegraphics[width=0.9\textwidth]{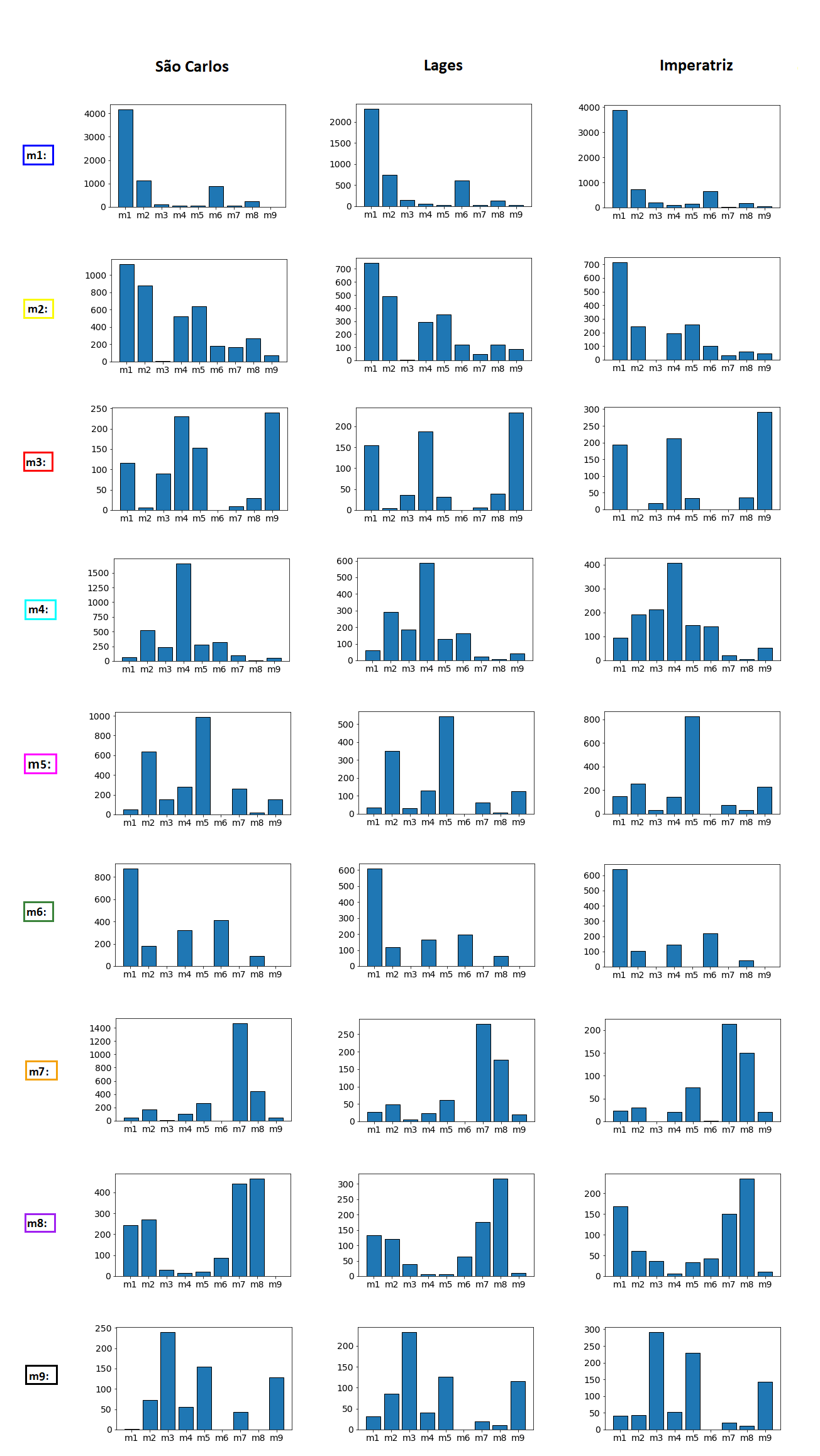}
    \caption{Histograms of the geographical adjacency between the nine identified motifs respectively to the cities of S{\~a}o Carlos, Lages, and Imperatriz.  These results corroborate the tendency of certain motif types presenting higher probability of being near one another.}
    \label{fig:neighborFreq}
\end{figure}

Indeed, it could be expected that some types of motifs tend to appear adjacent one another as one moves from more uniform to less uniform, or from more central to more periphery regions of a city.  In order to verify this possibility in an objective and quantitative manner, Figure~\ref{fig:neighborFreq} depicts the histograms of city motif adjacencies in the cities of S{\~a}o Carlos (a), Lages (b) and Imperatriz (c).

\begin{figure}[ht!]
    \centering
    \includegraphics[width=\textwidth]{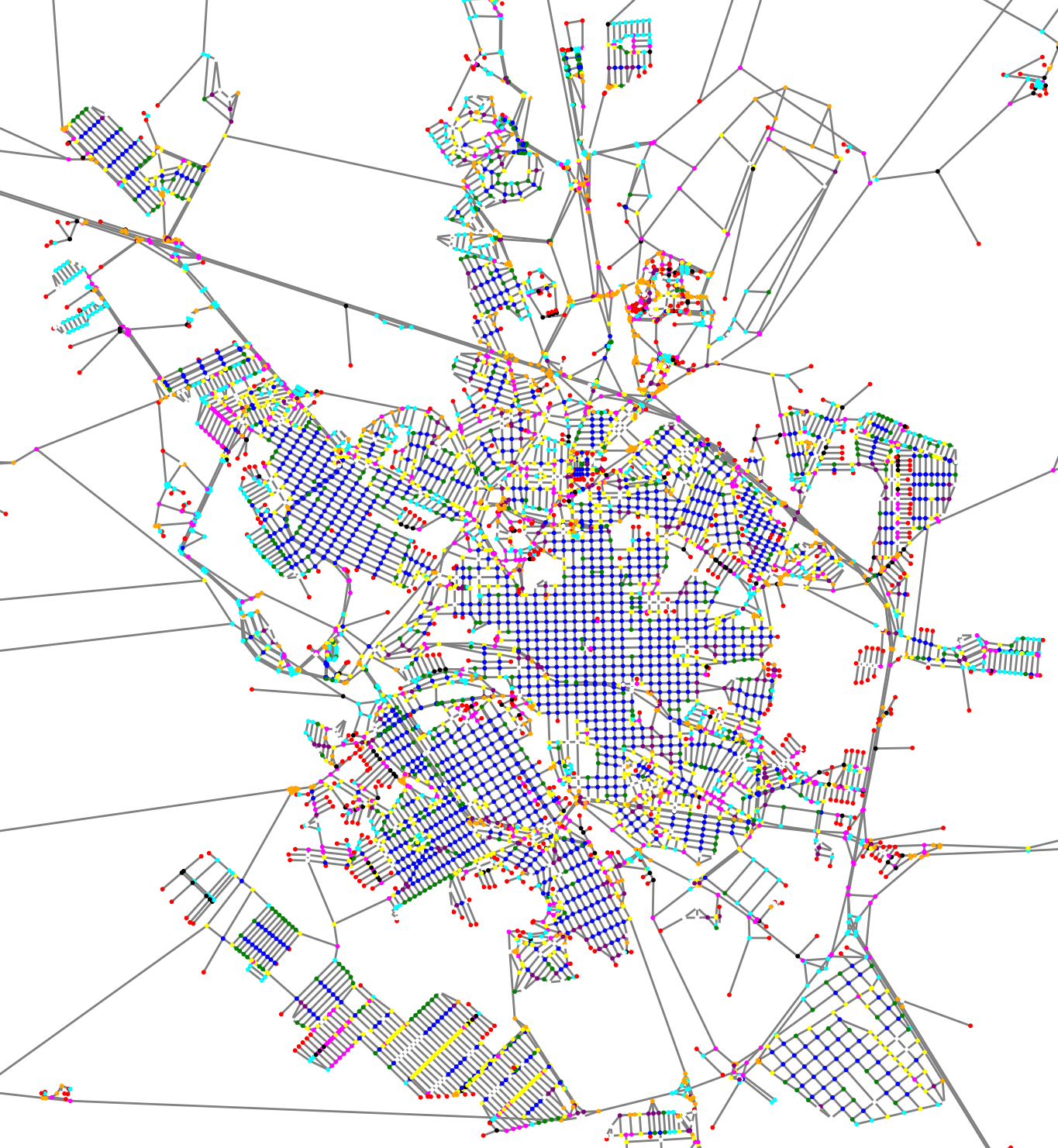} 
    \caption{The geographical distribution of motif types in the city of S{\~a}o Carlos.}
    \label{fig:geoSaoCarlos}
\end{figure}

\begin{figure}[ht!]
    \centering
    \includegraphics[width=\textwidth]{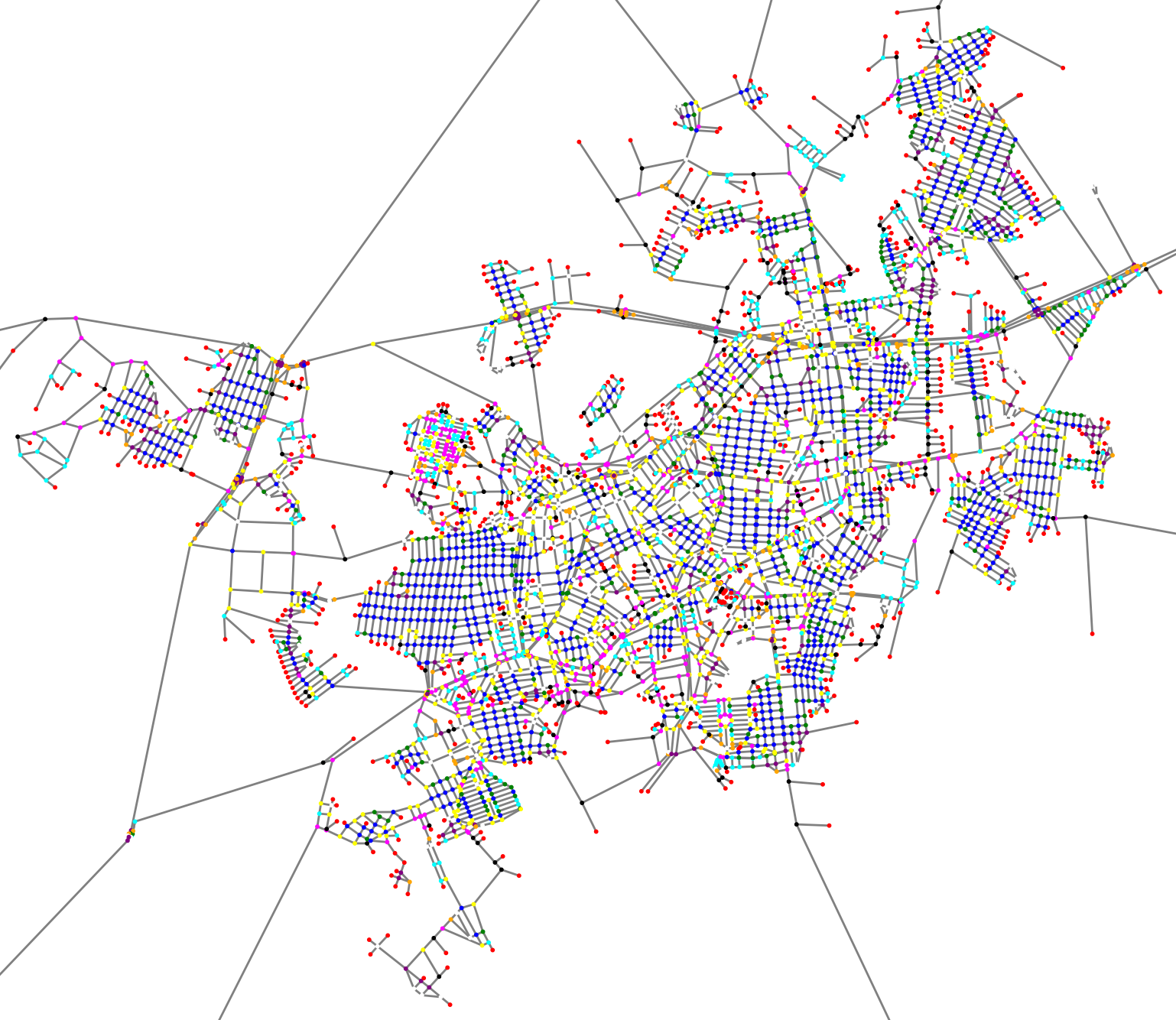} 
    \caption{The geographical distribution of motif types within the city of Lages.}
    \label{fig:geoLages}
\end{figure}

\begin{figure}[ht!]
    \centering
    \includegraphics[width=\textwidth]{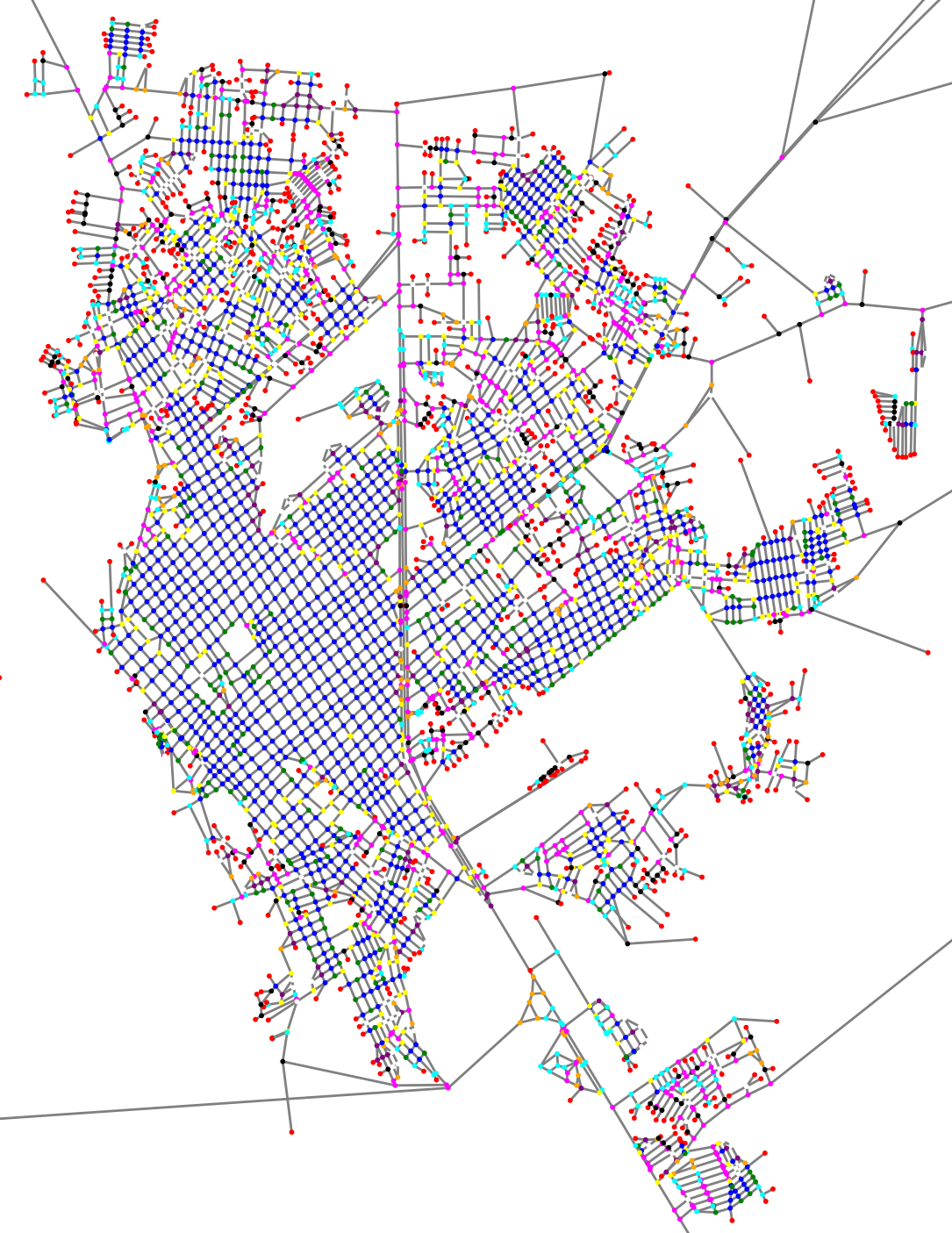} 
    \caption{The geographical distribution of motif types within the city of Imperatriz.}
    \label{fig:geoImperatriz}
\end{figure}

The spatial relationships between motif types can be more effectively visualized in terms of their distribution within each of the considered cities, as shown in Figures~\ref{fig:geoSaoCarlos},~\ref{fig:geoLages}, and~\ref{fig:geoImperatriz}. 

It is interesting to keep in mind that, given a neighborhood network, where each node corresponds to the reference node of the respective neighborhood, the fact that two nodes $i$ and $j$ are adjacent implies overlap between their respective neighborhoods.  As a consequence, two adjacent neighborhoods tend to have similar local topological properties.  That is one of the reasons why each of the motif types tends to present specific adjacency preferences.

The results in Figure~\ref{fig:neighborFreq} corroborate the tendency of the motifs types to appear adjacent one another in specific manners.  A first interesting result that can be identified in this figure regards the fact that the motifs adjacency tends to be consistent among the three considered cities.   

Regarding the predominant adjacencies observed for the nine identified motifs, we have that the motif types $m1$, $m2$ and $m6$ tend to be adjacent to themselves.  This transitive property is of particular importance as it leads to patches of neighborhoods sharing the same motif type.  For instance, motif $m1$ tends to form extensive regions of almost perfect orthogonality, and therefore regularity, in cities.  Interestingly, this type of motif tends to be adjacent to $m2$ or $m6$, frequently appearing at the border of the regular patches identified as $m1$.  Motif $m3$ is mostly adjacent to motif type and $m9$.  Given that $m3$ often corresponds to streets dead-ends, we also have that the motif type $m9$ also tends to occur near the geographical borders between the communities within cities. 
Motif type $m7$ presents a preferential tendency to be adjacent to motif $m8$, being related to triangular blocks.

It should be observed that having similar topological properties contributes to making a pair of motif types to appear geographically adjacency, but this is not always the case.  Take, for instance, motif types $m7$ and $m9$ in Figure~\ref{fig:neighborFreq}.  Though the surroundings of their reference nodes both tend to present a radiating tree-like structure in both cases, therefore sharing several hierarchical properties as indicated by the strong respective connection in Fig.~\ref{fig:redeMotif}), the reference node in $m7$ often corresponds to one of the vertices of a triangular block, which is not the case of $m9$.  For this reason, though topologically similar, these two motif types are highly unlikely to be found geographically adjacent in a city.

\section{The Nine Identified Motifs} \label{sec:motifs}

By referring to Figures~\ref{fig:redeMotif}, ~\ref{fig:motifs}, ~\ref{fig:hist}, and~\ref{fig:neighborFreq}, we can now typify each of the nine identified motifs as follows: \newline

\noindent \emph{\textbf{$m1$, Blue:}} As it can be observed in Figure \ref{fig:motifs}, this motif type tends to have its reference node with degree 4.  In addition, we have from Figure~\ref{fig:hist} that this motif is characterized (together with $m8$) by the highest hierarchical degree and hierarchical number of nodes, as well as for particularly high values of the $hn$.   As it can be discerned from Figure~\ref{fig:motifs} as well as the geographical distributions (Figures~\ref{fig:geoSaoCarlos} --~\ref{fig:geoImperatriz}), that this motif is intrinsically associated to highly regular patches of square blocks.  This motif type also tends to appear adjacent to itself as well as to $m2$ and $m6$.
\newline

\noindent \emph{\textbf{$m2$, Yellow:}} This motif type, whose reference nodes tend to have degree 3, is similar to $m5$ and $m6$.  However, $m2$ and $m5$ have the $hd$ histograms at different positions.  At the same time, the $cr$ densities are at different positions in $m2$ and $m6$, and the latter motif has a wider dispersion of $hd$. Motf $m2$ tends to appear adjacent to itself, $m1$, and $m5$.  This motif, which tends to have a relatively large number of second neighbors, ofter corresponds to irregular neighborhoods internally to patches of $m1$ motifs.  \newline 

\noindent \emph{\textbf{$m3$, Red:}} This motif type tends to have reference node with degree 1 or 2.  In addition, we have from Figure \ref{fig:hist} that this motif has the smallest hierarchical degree and hierarchical number of nodes.   Unlike $m1$, this motif does not tend to appear adjacent to itself, having a predominance to have adjacency with $m4$ and $m9$.  Figures \ref{fig:geoSaoCarlos} -- \ref{fig:geoImperatriz} indicates that this motif type tend to correspond to streets dead-ends, being therefore expected to appear mostly near the city borders.  \newline

\noindent \emph{\textbf{$m4$, Cyan:}} This motif, which often has reference node with degree 3, is similar to $m5$, but it tends to have $cr$ larger than that of $m5$.  In addition, $m4$ has a wider dispersion of $hd$.  Interestingly, this motif appears adjacent mostly to itself, and then with $m2$.  \newline

\noindent \emph{\textbf{$m5$, Magenta:}} This motif, with reference node tending to have degree 3, is similar to motifs $m2$, $m4$, and $m9$.  However, the $hd$ histograms are different among these three motifs.  In particular, $m9$ tends to have reference nodes with the smallest degree among these motifs.  This motif tends to be adjacent to itself and to $m2$. Generally speaking, motifs $m2$, $m4$, and $m5$ are typically found at the interfaces or transitions between the highly regular patches of $m1$ motifs.  \newline

\noindent \emph{\textbf{$m6$, Green:}} The reference node associated to this type of motif tends to have node degree equal to 3.  Its hierarchical measurements are largely similar to those of $m2$, though presenting $cr$ smaller.  This motif type tends to be predominantly adjacent to $m1$ and itself.  Figures \ref{fig:geoSaoCarlos} -- \ref{fig:geoImperatriz} indicate that this type of motif tends to correspond to borders of the highly regular patches of $m1$ motifs.  \newline

\noindent \emph{\textbf{$m7$, Orange:}} This motif type tends to have reference characterized by node degree equal to 3, as well as by relatively low $hd$values.   As it can be readily inferred from Figures \ref{fig:geoSaoCarlos} -- \ref{fig:geoImperatriz}, this motif type is characterized by having its reference node as corresponding to one of the vertices of a triangular block.  Motif type $m5$ tends to be adjacent to itself as well as to $m8$.  \newline

\noindent \emph{\textbf{$m8$, Purple:}} This is the second least frequently observed type of motif, with only 909 occurrences among the three considered cities.  The node degree of its reference node tends to vary between 4, and 5, as well as relatively high average value and dispersion of $hd$ values.  It is most similar to $m1$, but the latter has larger $hd$.  As motif type $m7$ the reference node of $m8$ tends to correspond to one of the vertices of a triangular block.  Motif type $m8$ tends to be adjacent to itself and $m7$. \newline

\noindent \emph{\textbf{$m9$, Black:}} The reference node of this motif type tends to have degree equal to 3.  It is most similar to $m5$ and $m7$.  However, $m9$ tends to have $he$ smaller than $m7$, and $hd$ distinct $m5$.  This type of motif tends to be adjacent to $m3$, itself, and $m5$. Together with nodes associated to motif $m3$ motif $m9$ plays an important role in helping to identify the contours of the cities.\newline

\section{Analysis of the Influence of the Adopted Features}

Almost invariably, the results obtained from comparisons and classifications depend substantially on the adopted measurements or features used to characterize each data element.  Even though the five selected features (Section~\ref{subsec:hierarchical}) allowed remarkable results regarding the identification of city motifs, it is still interesting to study the effect of each of the adopted feature on the obtained motif networks.  The present section focuses on this aspect.

In order to do so, in the present work we apply the feature analysis methodology described in~\cite{costa2021elementary,costa2022cities}.  More specifically, coincidence networks are obtained considering all possible combinations of the adopted features. Each of these networks is represented by the respective weight matrix, whose entries correspond to the obtained coincidence values. Then, the coincidence similarities are obtained between every pair of respective weight matrices, yielding a \emph{features network}.  Each node in the latter corresponds to a coincidence network respective to some features combination, while the link weights indicate the respective coincidence similarities.

\begin{figure}[ht!]
    \centering
    \includegraphics[width=.7\textwidth]{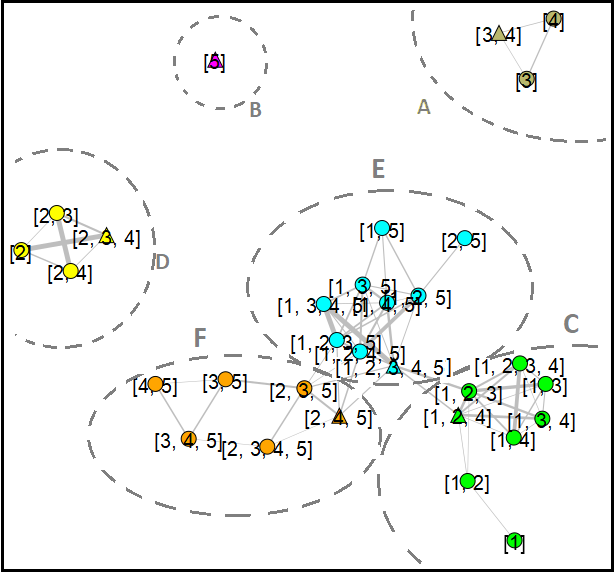} \vspace{0.5cm}
    \caption{The \emph{features network} obtained for the motifs identified for the city of Lages.  Each node corresponds to one of the 31 possible combinations of the five adopted features (hierarchical measurements), while the links width is proportional to the value of the coincidence similarity between the motifs networks obtained for the respective combinations and assuming $T = 0.3$ and $\alpha = 0.1$.  Six communities, identified by respective color, have been identified, which can be understood as the main possible models of the effect of the features on the obtained motifs network.  The hubs within each identified community are shown as triangles.  The colors in this figure were used only for highlighting the six models, bearing no relationship whatsoever with the identified motifs.}
    \label{fig:att}
\end{figure}

In the present work, we will focus on deriving and discussing the features vector respective to the city of Lages, which has a good balance among the identified motifs.  Given that five measurements (features) have been adopted, corresponding to hierarchical measurements, the resulting features network will necessarily have 31 nodes, each corresponding to a possible combination, except for the null case.   Figure~\ref{fig:att} depicts the therefore obtained features network.

A total of six communities have been found by using the infomap methodology (e.g.~\cite{rosvall2008maps}), each of which corresponding to a respective \emph{putative model} of the motifs networks that can be obtained for different features combinations.  Interestingly, the network obtained while considering all the five features resulted right at the center of the obtained network, being strongly interconnected to other nodes.

\begin{figure}[ht!]
    \centering
    \includegraphics[width=\textwidth]{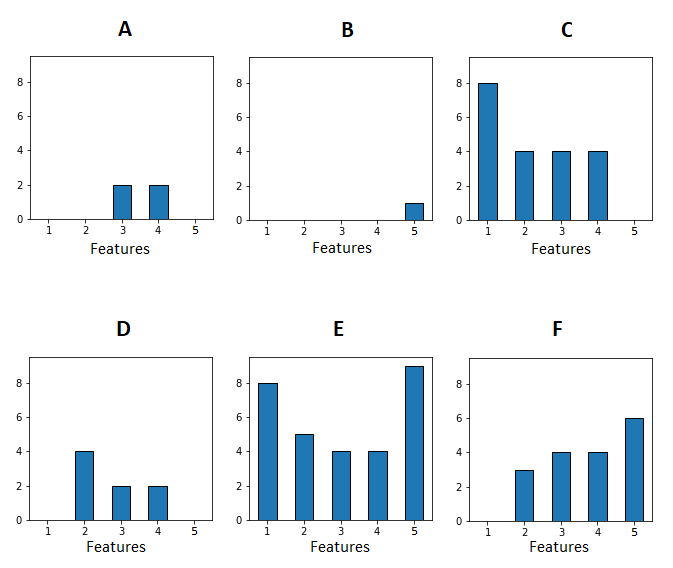}
    \caption{Histograms of the features adopted within each of the six obtained models describing the effect of features combinations, respectively to the communities A -- F.}
    \label{fig:histatt}
\end{figure}

Further understanding of the influence of the features can be derived by taking into account the histograms of features to be found within each detected community.  These histograms are shown in Figure~\ref{fig:histatt}.

The motifs in model C employ the features 1 to 4, while the feature 5 is not found in this model.  The model D involves features 2 to 4.  All features contribute to the motifs networks in model E, with predominance of the features 1 and 5.  

All in all, we have that the adopted features lead to six main putative models of the networks, with the model E corresponding to being the most interconnected within itself as well as with the other models (more central), having special relevance.

\section{A Simple Supervised Method for Assigning Motifs}

Given that the nine identified city motifs depend exclusively on local measurements, namely only the two neighborhood levels around each reference node, they tend not to be influenced by the remainder of the streets networks and be relatively immune to border effects.  In addition, it is arguable that the local city topology is largely universal as it is required to cater for similar demands, such as transportation, mobility, access to resources, etc.  Yet another important aspect possibly supporting the generality of the identified motifs is the fact that streets networks are largely geographical networks with scant long range connections.   

In the light of the above discussion, it is reasonable to posit that the identified motifs can be relatively universal.  Under this assumption, it becomes possible to transfer the learned motifs to other cities, which can be done in a remarkably simple manner.  First, some cities are taken as models, and their combined neighborhood network is respectively obtained as described in the current work.  Then, a table is derived in which each line corresponds to one of the neighborhoods of the combined network that have been identified as motifs, followed by its respective five hierarchical measurements.  Now, given a neighborhood from any other city to be classified, its features can be compared to those in the reference table and, in case the maximum coincidence is larger than a given threshold, the motif type of the respective entry in the table is assigned to the new neighborhood.  

Observe that, while the motif identification approach described previously in the current work can be understood as being unsupervised, the table-based method is supervised.

It is interesting to observe that the above suggested methodology can also be applied in case the cities to have their motifs identified come from a same set of cities that are known to share their topological properties.  In other words, the reference motifs do not need to be completely universal among all possible cities, but only within a given set of cities with similar topology.

\begin{figure}[th!]
    \centering
    \includegraphics[width=\textwidth]{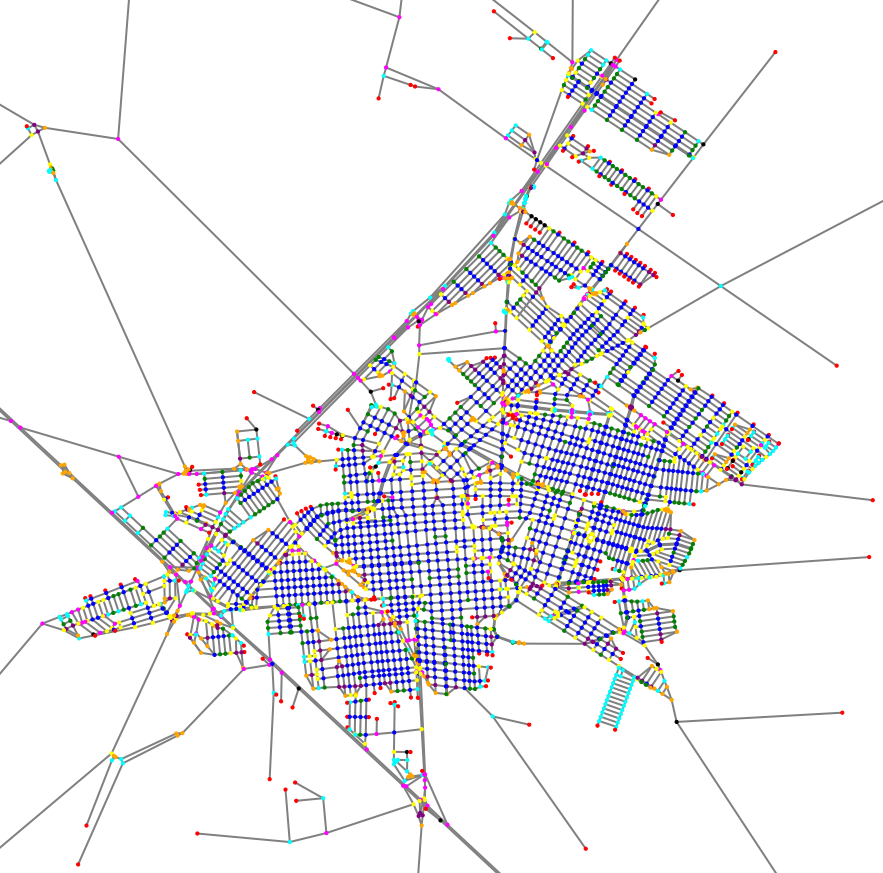}
    \caption{The motifs associated to the neighborhoods of Birigui by using the suggested motif assignment method while considering the motif reference table obtained for the cities of S{\~a}o Carlos, Lages and Imperatriz taken as reference for motif identification.}
    \label{fig:SC}
\end{figure}

Figure~\ref{fig:SC} presents an example of result obtained by the application of the above described methodology with respect to the new Brazilian city of Birigui.   The reference table obtained for the three cities considered in the present work was used for the supervised classification of the neighborhoods of this new city.

The obtained result indicates that most motifs have been properly identified, which corroborates the possible universality of the nine identified motifs across distinct cities.  This result motivates the application of the above described simple supervised network to other cities, paving the way to many possible further analysis of urban organization.

Table~\ref{tab:freqBirigui}  presents the relative frequency of the motifs obtained for the city of Birigui.  It could be expected that the frequencies characterizing distinct cities result distinct, reflecting intrinsic geographical and other types of environmental specificities.  This can be observed by comparing the obtained relative frequencies in Tables~\ref{tab:freqBirigui} and~\ref{tab:motifs}. We have a markedly large number of motifs $m1$, suggesting a more predominant orthogonal organization of this city. In addition, a relatively smaller frequency of motifs of types $m3$, $m5$ and $m9$, while motif $m6$ tended to appear more frequently.  

\begin{table}[]
    \centering
    \begin{tabular}{c|c|c|c}
        Motif identif. & Motif color & N. of nodes & Rel. Freq.\\ 
        \hline  
        $m1$ & blue & 4015 & 36.038 \% \\
        $m2$ & yellow & 1826 & 16.411 \% \\ 
        $m3$ & red & 804 & 7.226 \% \\ 
        $m4$ & cyan & 1041 & 9.356 \% \\ 
        $m5$ & magenta & 755 & 6.785 \% \\ 
        $m6$ & green & 1123 & 10.093 \% \\ 
        $m7$ & orange & 858 & 7.711 \% \\ 
        $m8$ & purple & 571 & 5.132 \% \\
        $m9$ & black & 134 & 1.204 \% 
    \end{tabular}
    \caption{The identification and number of neighborhoods assigned to each of the nine identified motifs obtained for Birigui.}
    \label{tab:freqBirigui}
\end{table}

\section{Concluding Remarks}

The study and characterization of cities has constituted the focus of significant attention along the last decades, especially given the potential of such analysis for contributing to sharing administrative experiences, enhancing urban aspects, and better understanding relationships between the city topology and other socioeconomic factors, among several others possibilities.

In network science, the concept of network motifs has been applied with particular effectiveness for characterizing and better understanding the network topology.  Here, we approached the interesting topic of city characterization in terms of statistical motifs identified from network representations of cities.  More specifically, we adopted a local characterization of the topological features of neighborhoods around the respective nodes.  This has been accomplished by using five hierarchical measurements considering two neighborhood levels around each reference node.  The pairwise similarity between these neighboorhods was then estimated by using the coincidence methodology, which implements a particularly strict similarity quantification contributing to higher levels of connectivity detail and network modularity.  Neighborhood networks, obtained for three Brazilian cities from distinct regions and with distinct topological characteristics, were then combined into a single network, which had its communities detected by the Infomap. The nine city motifs were therefore identified were remarkably consistent not only within a same city, but also across the three considered cities, suggesting that they may have a universal comprehensiveness.  This potentially remarkable result is supported by the locality of the adopted measurements, which are limited to two hierarchical levels around the reference node. 

The properties of the identified motifs were then characterized and discussed based on four main perspectives, namely the motifs similarities, visualizations of samples of each motif, distributions of the five adopted hierarchical measurements, as well as histograms of adjacency between the nine motifs.  The obtained city motifs can be understood from both the perspective of homogeneity, complexity, as well as centrality, with one of the motifs ($m1$) corresponding to the prototypical square block organization characterizing full orthogonal street plans.  This type of motif tends to be the most regular and central among the identified types.  Other particularly interesting motifs, $m3$ and $m9$, tend to appear near streets dead ends, being therefore found predominantly along the city contour.  Motifs $m7$ and $m8$ both have their reference nodes corresponding to one of the vertices of a triangular block, but they distinguish one another respectively to other hierarchical measurements. Motifs of type $m2$, $m4$ and $m5$ tended to be particularly irregular, frequently appearing as an interface or transitions between more regular patches.  These two motifs, however, have distinct hierarchical degrees.  

As a complement to the reported approach to city motifs identification, we also performed an analysis of the influence of the adopted hierarchical features on the respectively obtained neighborhood networks.  This was accomplished by using the coincidence similarity, leading to the identification of six possible models (communities) of neighborhood networks that can be obtained by combining the five adopted features.  The most cohesive model involves all the five adopted hierarchical measurements, with the hierarchical degree and convergence ratio predominating in this model.  

Although the proposed methodology to identify city motifs involves several concepts and steps, a simple supervised method has been also suggested and illustrated in this work for assigning motif types to a given streets network.  This procedure is based on a reference table containing several instances of neighborhoods and their motifs identified respective to a set of reference cities used for training.  Then, given a new city represented in terms of the respective streets networks, motif types can be assigned to its neighborhoods by taking into account the motif of the table entry presenting the features that are more similar to those of each of the new nodes.  The potential of this simple supervised methodology has been illustrated with respect to the Brazilian city of S\~ao Carlos with remarkable results.  This simple method for motif assignment assumes substantial level of the universality of the identified motifs, which constitutes an aspect to be further substantiated.

The encouraging results reported in the present work respectively to concepts, methodology and results, pave the way for a large number of future possible developments. For instance, it would be interesting to investigate the effect of larger neighborhood extensions ($H$) on the resulting motifs.  It would also be interesting to compare cities based on their respective distribution of motifs, as well as the adjacency between them.  In particular, the patches indexed by the same type of motifs can be easily identified (e.g.~by using connected component methods) so that their topological and geometrical properties can be studied at a spatial scale larger than the adopted $H$.  
Given the inherently hierarchical nature of the accessibility (e.g.~\cite{travenccolo2008accessibility,de2014role,viana2013accessibility}), it would also be of interest verifying the respectively implied motifs when adopted instead (or as a complement) of the hierarchical measurements.    Another particularly promising perspective regards the incorporation of geometrical features as a means to complement the topological features adopted in the present work.  For instance, even more strict identification of motifs belonging to highly orthogonal portions of a city can be obtained by taking into account also the lengths of each of the block sides.

In addition, given that motifs can be expected in a wide range of real-world and theoretical networks, its would be of great interest to apply the concepts and methodology proposed in the present work to other types of networks, such as roads and airport routes, energy distribution, Internet and WWW, protein interaction, scientific collaboration, text and citations networks, among many other possibilities.

\section*{Acknowledgements}

G. S. Domingues thanks CAPES (88887.601529/2021-00) for financial support. 
E. K. Tokuda thanks FAPESP (2019/01077-3) for financial support. 
L. da F. Costa thanks CNPq (307085/2018-0 ) and FAPESP (215/22308-2) for support.
This study was financed in part by the Coordenação de Aperfeiçoamento de Pessoal de Nível Superior – Brasil (CAPES) – Finance Code 001.

\bibliography{main}
\bibliographystyle{unsrt}

\end{document}